\documentclass[useAMS,usenatbib]{mn2e}
\usepackage{graphicx}
\usepackage{soul}

\title[Cepheids observed by \textit{MOST}]{V473 Lyr, a modulated, period-doubled Cepheid, and U~TrA, a double-mode Cepheid observed by \textit{MOST}\thanks{Based on data from MOST (Microvariability \& Oscillations of STars), which was, at the time the data reported here were collected, a Canadian Space Agency mission operated jointly by Microsatellite Systems Canada Inc. (MSCI, formerly the Space Division of Dynacon Inc.), and the Universities of Toronto and British Columbia, with support from the University of Vienna.}}
\author[L. Moln\'ar et~al.]{L.\ Moln\'ar$^{1}$\thanks{E-mail:
molnar.laszlo@csfk.mta.hu}, A.\ Derekas$^{2,1}$, R.\ Szab\'o$^1$, J.\ M.\ Matthews$^{3}$, C.\ Cameron$^{4,5}$,
\and A.\ F.\ J.\ Moffat$^6$, N.\ D.\ Richardson$^7$, B.\ Cs\'ak$^{2}$, \'A.\ D\'ozsa$^{2}$, P.\ Reed$^8$, L.\ Szabados$^{1}$, 
\and B.\ Heathcote$^{9,10}$, T.\ Bohlsen$^{10}$, P.\ Cacella$^{10}$, P.\ Luckas$^{11}$, \'A.\ S\'odor$^{1}$, M.\ Skarka$^{1}$, 
\and Gy.\ M.\ Szab\'o$^2$, E.\ Plachy$^1$, J.\ Kov\'acs$^2$, N. R. Evans$^{12}$, K.\ Kolenberg$^{13,14,12}$, 
\and K.\ A.\ Collins$^{15,16}$, J.\ Pepper$^{17}$, K.\ G.\ Stassun$^{16,18}$, J.\ E.\ Rodriguez$^{16}$, R.~J.\ Siverd$^{19}$,  
\and  A.\ Henden$^{20}$, L.\ Mankiewicz$^{21}$, A.\ F.\ \.Zarnecki$^{22}$, A.~Cwiek$^{23}$, M.~Sokolowski$^{23,24}$, 
\and A. P\'al$^{1}$, D.\ B.\ Guenther$^{25}$, R.\ Kuschnig$^{26}$,  J.\ Rowe$^{5}$, S.\ M.\ Rucinski$^{27}$, 
\and D.\ Sasselov$^{11}$, W.\ W.\ Weiss$^{26}$\\} 

\begin{document}

\date{Accepted ... Received ...; in original form ...}

\pagerange{\pageref{firstpage}--\pageref{lastpage}} \pubyear{2016}

\maketitle

\label{firstpage}

\begin{abstract}
Space-based photometric measurements first revealed low-amplitude irregularities in the pulsations of Cepheid stars, but their origins and how commonly they occur remain uncertain. To investigate this phenomenon, we present \textit{MOST} space telescope photometry of two Cepheids. V473 Lyrae is a second-overtone, strongly modulated Cepheid, while U Trianguli Australis is a Cepheid pulsating simultaneously in the fundamental mode and first overtone. The nearly continuous, high-precision photometry reveals alternations in the amplitudes of cycles in V473 Lyr, the first case of period doubling detected in a classical Cepheid. In U TrA, we tentatively identify one peak as the $f_X$ or 0.61--type mode often seen in conjunction with the first radial overtone in Cepheids, but given the short length of the data, we cannot rule out that it is a combination peak instead. 

Ground-based photometry and spectroscopy were obtained to follow two modulation cycles in V473 Lyr and to better specify its physical parameters. The simultaneous data yield the phase lag parameter (the phase difference between maxima in luminosity and radial velocity) of a second-overtone Cepheid for the first time. We find no evidence for a period change in U TrA or an energy exchange between the fundamental mode and the first overtone during the last 50 years, contrary to earlier indications.

Period doubling in V473 Lyr provides a strong argument that mode interactions do occur in some Cepheids and we may hypothesise that it could be behind the amplitude modulation, as recently proposed for Blazhko RR Lyrae stars. 

\end{abstract}


\section{Introduction}

Classical Cepheids, also known as $\delta$ Cep-type stars or just Cepheids, are Population I Instability Strip pulsators crucial for understanding stellar structure and evolution. Most Galactic Cepheids pulsate in one mode with remarkably stable amplitudes and periods, although space-based photometry reveals that they are not entirely regular clocks. Observations with \textit{CoRoT, Kepler} and \textit{MOST} detect short-term, low-amplitude irregular variability or "jitter" in the pulsation cycles of some classical Cepheids \citep{derekas2012,evans2015,poretti2015}.  

One exception is V473 Lyrae ($\alpha_{2000.0}$~=19$^{\rm h}$15$^{\rm m}$59\fs49, $\delta_{2000.0}$ = +27\degr 55\arcmin 34\farcs 6, $V$ = 6.18 mag), the only known classical Cepheid in the Galaxy that undergoes strong amplitude and phase modulations. The changes in the pulsation amplitude were first noted by \citet{bm80}. \citet{pe80} estimated the modulation period to be 3.3--3.4 years, which was later measured to be 1210 d by  \citet{breger81}, and 1258 d by \citet{cab91}, respectively. The origin of this modulation was extensively discussed in the literature, with theories ranging from mode beating \citep{breger} to magneto-convective cycles \citep{stothers}. A recent analysis of extensive ground-based observations collected during 1966--2011 \citep{molnar2014} confirmed that V473 Lyrae pulsates in the second overtone, modulated with a primary period of $1205\pm3$ d. \citet{molnar2014} also discovered a second, longer modulation cycle with a period of $5300\pm150$ d, and concluded that the modulation closely resembles the Blazhko effect seen in RR Lyrae stars. V473 Lyrae is unique among Galactic Cepheids: the closest counterparts are V1154 Cyg, in which the \textit{Kepler} space telescope detected modulation but with a very low amplitude \citep{kanev2015, derekas2016} and SV Vul, with a suspected, much longer modulation period of order 30 yr \citep{engle2014}. Beyond our Galaxy, strong amplitude modulation was recently found in three classical Cepheids in the Magellanic Clouds \citep{soszynski2015}. 

Although modulation appears to be very rare among Cepheids, the phenomenological properties of the light variations of V473 Lyr closely resemble the Blazhko effect, a common feature among RR Lyrae stars. Thanks to space-based photometry, we now know that many modulated RR Lyrae stars show low-amplitude additional modes and/or alternating high- and low-amplitude cycles known as period doubling \citep{benko2010,szabo2010,szabo2014}. Numerical models suggest that mode resonances that generate the period doubling can also be responsible for the modulation itself \citep{bk11}. It is not unreasonable, therefore, to postulate that similar mechanisms could operate in modulated Cepheids too. 

Unfortunately, the pulsation period of V473 Lyr, $P$~=~1.4909\,d, means that one cannot observe consecutive pulsation maxima from a single ground-based site. This makes it virtually impossible to detect period doubling, a challenge also faced in ground-based RR Lyrae observations. Therefore, we proposed to gather continuous observations of the star with the \textit{MOST} space telescope for a timespan of one month in order to search for cycle-to-cycle variations. 

\textit{MOST} serendipitously also observed another Cepheid, U TrA, just before the run that targeted V473 Lyr. A campaign aimed at the Wolf-Rayet star WR 71 included this beat Cepheid within the same field of view. U TrA ($\alpha_{2000.0}$ =16$^{\rm h}$07$^{\rm m}$19\fs01, $\delta_{2000.0}$ = --62\degr 54\arcmin 38\farcs 0, $V$ = 7.89 mag), along with TU Cas were the first Cepheids in which beating (the characteristic feature caused by double-mode pulsation) was observed \citep{oo57a,oo57b}. U TrA pulsates in the fundamental mode and the first overtone, with periods $P_0 = 2.56842$\,d and $P_1 = 1.82487$\,d \citep{pardoporetti}. After an investigation of the relative amplitudes of the pulsation modes over time, \citet{faulknershobbrook79} proposed that pulsation energy may have been redistributed from the fundamental mode to the first overtone, an intriguing concept that can be tested with new measurements. Together with earlier observations of RT Aur and SZ Tau \citep{evans2015}, \textit{MOST}'s Cepheid sample now totals 4 stars.

This paper is structured as follows: Section \ref{sec2} describes the various observations we carried out; Sections \ref{secv473} and \ref{secutra} include the analysis and obtained results for V473 Lyr and U TrA, respectively; Section \ref{secsum} closes with discussion and summary of the findings presented in the paper.

\begin{figure*}
\includegraphics[width=1.0\textwidth]{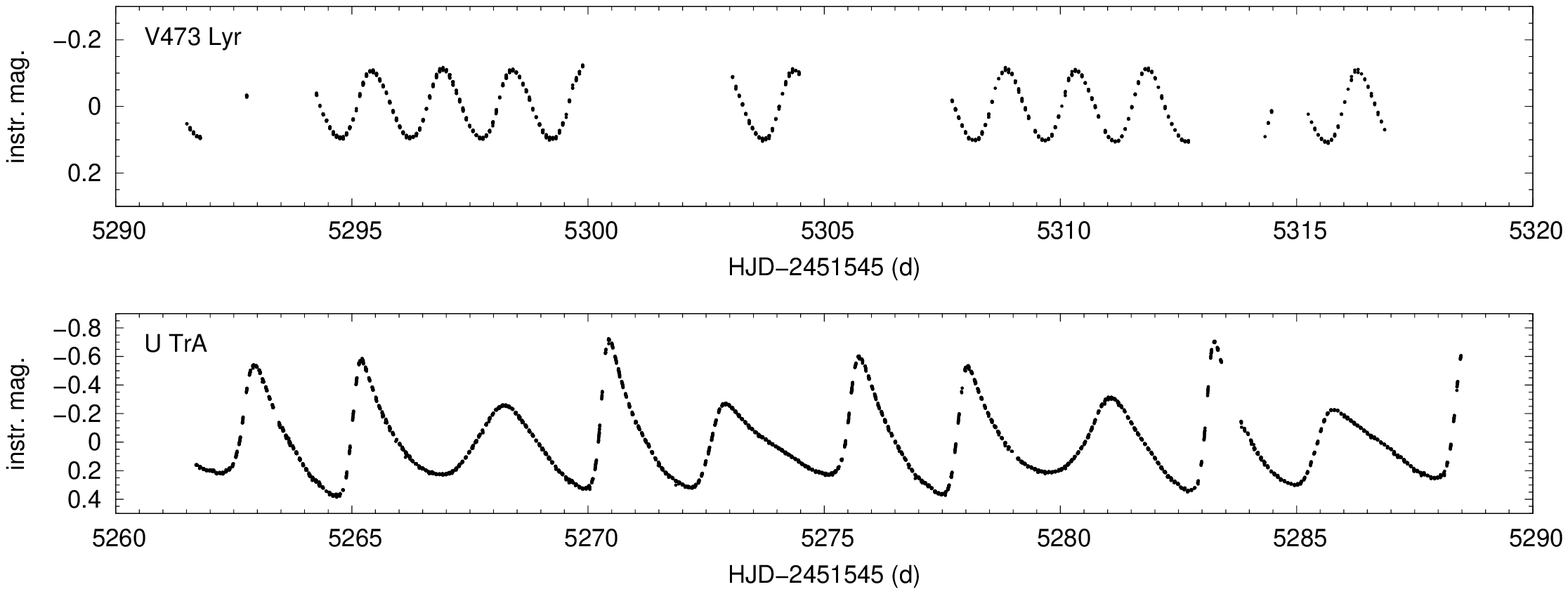}
\caption{\textit{MOST} light curves of V473 Lyr and U TrA. Brightnesses are in \textit{MOST} instrumental units. }
\label{most-lc}
\end{figure*}

\section{Observations}
\label{sec2}
The stars were observed by the \textit{MOST} (Microvariability \& Oscillations of STars) space telescope: a Canadian microsatellite designed to observe bright stellar targets continuously for up to 2 months at a time \citep{walker03}. We also obtained spectroscopy for both stars with various ground-based instruments.

\subsection{\textit{MOST} observations and data reduction}

Neither of the two stars was observed continuously during each 101-minute orbit of the \textit{MOST} satellite.  

In the case of U TrA, in the field of the \textit{MOST} Primary Science Target WR 71, each orbit was shared with another Primary Science field. U TrA was observed for 28 days during 29 May -- 25 June 2014. As \textit{MOST} has to use the science CCD for guiding purposes too, it takes short individual exposures that are stacked on-board to boost the stellar signals. The sampling cadence is 123.8 sec.

In the case of V473 Lyrae, the star is located in the sky where \textit{MOST} must point at too large an angle relative to the Sun, heating the satellite beyond the thermal limits for its battery. But the V473 Lyrae run coincided with the ``eclipse season" for \textit{MOST}, when the Sun skims briefly below the Earth horizon once per orbit for a few minutes. It was possible to point \textit{MOST} at V473 Lyrae for a few minutes per orbit, leading to an effective sampling cadence of 101 minutes. Each orbit contains 2--10 measurements, with a cadence of 62.2 sec. The star was observed for 27 days during 28 June -- 23 July 2014, but communication problems with the satellite and deliberate interruptions to observe other targets introduced some longer gaps. We have data from 57 per cent of the \textit{MOST} orbits in the overall run.

MOST photometric reduction processes are well established \citep[see, e.g.,][]{reegenetal2006,rowe2006} and have recently been successfully applied to other Cepheids \citep{evans2015}. \textit{MOST} data experience modulation of scattered Earthshine with the 101-min orbital period.  Also, as the instrument regains thermal equilibrium after switching from two widely-separated points in the sky, there are short-lived trends after switching from one field to another. Stray light modulation is filtered in the reduction process; extreme outliers were removed by sigma-clipping the flux values. In instances of substantially poor pointing, data were filtered by sigma-clipping outliers in the pointing positions and the measured flux values. 

To preserve the intrinsic Cepheid variations, slow trends were filtered from the data and clipping applied to the residuals. The filtered data were phased with the orbital period and averaged in bins to remove modulations caused by stray light. We note that V473 Lyr is much less affected by stray light than U TrA as it was only observed when \textit{MOST} was in the shadow of the Earth. Finally, the large-amplitude variation was restored to the data. Any remaining slow trends were then subtracted by a simple linear fit. The final light curves (times, magnitudes and errors) are shown in Fig.~\ref{most-lc}.

\begin{table*}
\caption{Journal of ground-based spectroscopic and photometric observations of V473 Lyrae. Time ranges are in HJD 2400000+ d. The KELT survey uses Kodak Wratten No.~8 red-pass filters \citep{pepper-kelt}. Aperture ratios and focal lengths are provided for the telephoto lens-based systems. } 
\begin{tabular}{lcccccc}
\hline
\noalign{\vskip 0.1cm}
Observer/survey   &  Method     &  Telescope    &   Detector    &  Parameters      & No.\ of obs. & Range \\
\hline
AAVSO BSM         &  Phot. &  Takahashi FS-60CB & SBIG ST-8XME  &  \textit{BVRI}   & 4$\times$127 & 55269.91--55896.60 \\
AAVSO Sonoita     &  Phot. &  C14 0.3 m             &   SBIG STL-1001E       &  \textit{BVRI}   & 4$\times$487 & 53556.81--54292.84 \\
KELT-N            &  Phot. &  Mamiya 645 f/1.9 80mm  & Apogee AP16E  &  red-pass filter &  2894        & 54725.65--56986.59 \\
Pi of the Sky (1) &  Phot. &  Canon EF f/1.2 85mm   & custom CCDs   &  white light     &  774         & 54004.50--54946.88 \\
Pi of the Sky (2) &  Phot. &  Canon EF f/1.2 85mm   & custom CCDs   &  \textit{R}      &  285         & 55697.80--56200.49 \\
GAO run 1    &  Spectr. & 0.5 m RC         & eShel echelle &  $R=11000$         &  681         & 56760.50--56874.59 \\
GAO run 2    &  Spectr. & 0.5 m RC         & eShel echelle &  $R=11000$         &  554         & 57179.42--57226.52 \\
Kutztown run 1 &  Spectr. & 0.6 m RC         & eShel echelle &  $R=11000$         &  393         & 56796.68--56879.74 \\
Kutztown run 2 &  Spectr. & 0.6 m RC         & eShel echelle &  $R=11000$         &  242         & 57183.61--57228.84 \\
Kutztown run 3 &  Spectr. & 0.6 m RC         & eShel echelle &  $R=11000$         &  59         & 57494.76--57499.88  \\
Piszk\'estet\H{o} &  Spectr. & 1.0 m RCC        & eShel echelle &  $R=11000$        &  216         & 56837.39--56845.57 \\\hline
\end{tabular}
\label{table:ground_phot_journal}
\end{table*}

\begin{figure*}
\includegraphics[width=1.0\textwidth]{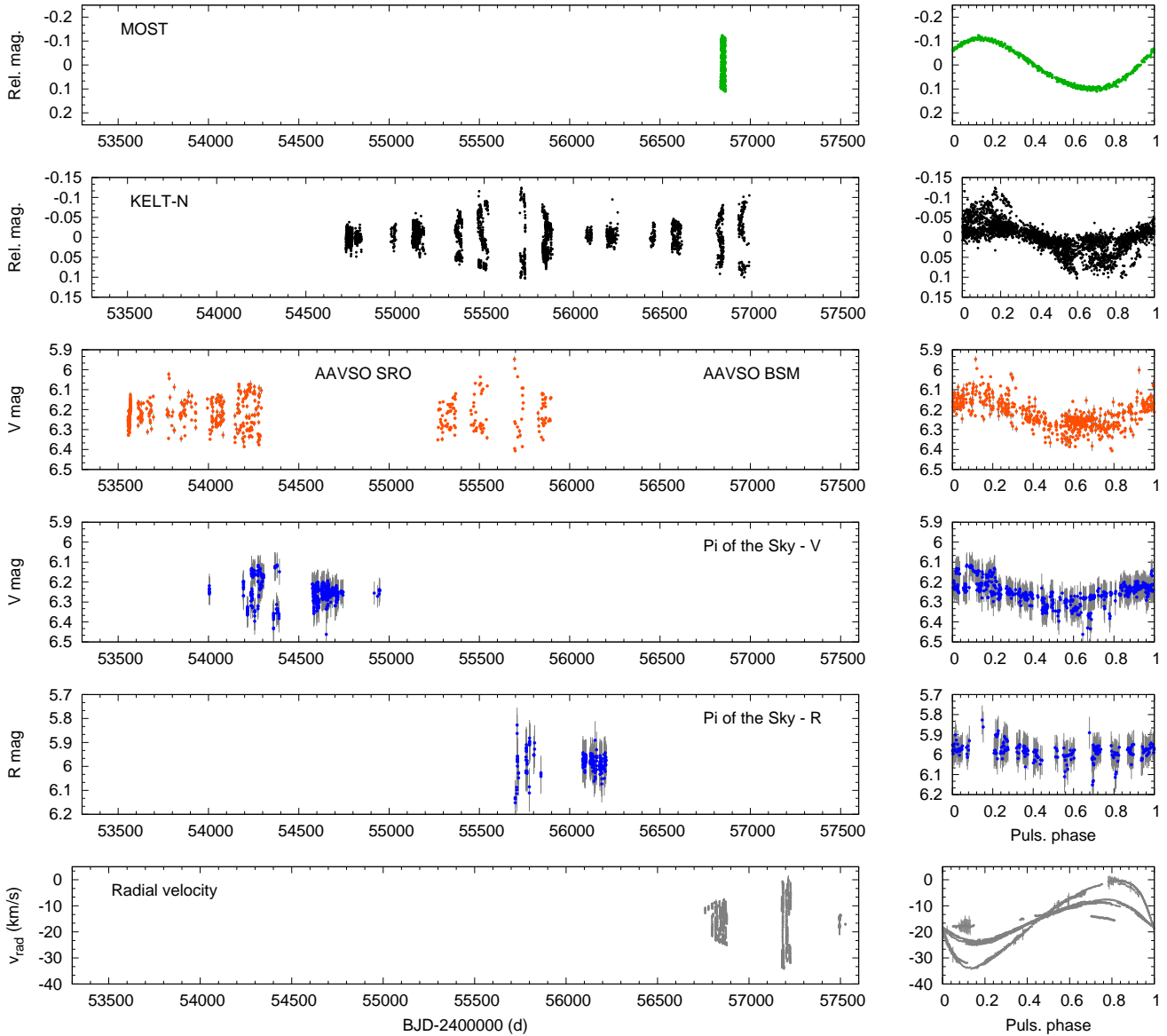}
\caption{New light curves and RV measurements of V473 Lyrae (left panels). Phased curves (right panels) are folded with the inverse frequency $P=1/0.670775$ d.  }
\label{v473_all_lc}
\end{figure*}

\subsection{Spectroscopic observations and reduction}

We organised an extended spectroscopic campaign to follow the radial velocity (RV) variations of V473 Lyrae, during 12 April -- 10 August 2014. We used a Shelyak Instruments \mbox{eShel} echelle spectrograph (resolving power of $ =11,000$ in the
spectral range $4200 - 8700$ \AA) mounted on the 0.5-m telescope of the Gothard Astronomical Observatory, Hungary (GAO), and for one week on the 1-m telescope at the Piszk\'estet\H{o} Mountain Station of the Konkoly Observatory (PO). For detailed descriptions of the device and the telescopes, see \citet{csak2014}. A similar spectrograph was used at Kutztown University Observatory, Pennsylvania, USA (KUO), mounted on a 0.6-m telescope.  The combined observations from Hungary and the USA extended the phase coverage, as the star returns to the same pulsation phase every third night owing to its near-1.5-day pulsation period. 

We repeated the campaign during 5 June -- 25 July 2015, with observations collected at GAO and KUO, and during 16 -- 21 April 2016 from KUO. All spectra were reduced using IRAF\footnote{IRAF is distributed by the National Optical Astronomy Observatories, which are operated by the Association of Universities for Research in Astronomy, Inc., under cooperative agreement with the National Science Foundation.} standard tasks, including bias and flat field corrections, aperture extraction, and wavelength calibration (using Th-Ar lines). We then normalised the continuum of the spectra. RV values were determined by cross-correlating the object spectra with R = 11,500 synthetic spectra chosen from the \citet{munari2005} library using the \texttt{fxcor} task of IRAF. Correlations were calculated between 4870--6550 \AA{} excluding Balmer lines, NaD and telluric regions. This method resulted in a 100--200 m\,s$^{-1}$ uncertainty in the
individual RV values. Barycentric Julian dates and velocity corrections for mid-exposures were calculated using the BARCOR code of \citet{hrudkova2006}. 

We also obtained two higher-resolution ($R=20,000$) spectra of V473~Lyr from PO in 2016. For those, barycentric correction was produced from IRAF. RVs were determined by the cross-correlation method using metal lines in the wavelength region 4800--5600~\AA. We fitted the cross-correlated line profiles with a Gaussian function to determine the RV values.

U TrA was observed spectroscopically by the members of SASER (Southern Astro Spectroscopy Email Ring)\footnote{http://saser.wholemeal.co.nz}, a group of advanced amateur and professional astronomers equipped with various instruments, with contributions from other observers as well. They collected 4 spectra in 2014 and 10 more in 2015. The observations include 3 low-resolution (R $\sim$ 1000--1500) spectra covering a wide wavelength range (3199--7199 \AA), and 11 higher-resolution spectra (R $\sim$ 4000--13,000) spanning a few hundred \AA. Radial velocities were determined with the task \texttt{fxcor} in IRAF, applying the cross-correlation method using a well-matching theoretical template spectrum from the extensive spectral library of \citet{munari2005}. 

Since the observed spectra cover different wavelength ranges, we determined velocities from each spectrum individually using their own wavelength region as listed in Table \ref{utra-sp-log-tbl}. We then made barycentric corrections to every RV value. However, we could not determine velocities from every spectrum, either because of the low resolution or the low number of lines in the observed wavelength region. Depending on the S/N and resolution of the spectrum, the estimated RV uncertainty ranges from 2 to 5 km/s. Derived values are listed in Table \ref{utra-sp-log-tbl} and discussed in Section \ref{sec-utra-rv}.

\subsection{Additional photometric data}

The photometric campaign for V473 Lyr by \citet{molnar2014} ended well before the \textit{MOST} run. We did not initiate a new photometric campaign, but searched databases of various surveys for data we might use. Unfortunately V473 Lyrae is too bright for most instruments used for exoplanet and asteroid searches. However, we managed to find data from 3 sources, mostly from various sky surveys that utilize various CCD cameras attached to telephoto lenses. 

The American Association of Variable Star Observers (AAVSO) provided us with \textit{BVRI} observations taken with their Bright Star Monitor instrument and the 0.5 m telescope at Sonoita Research Observatory, Arizona (USA). We have received data from the Pi of the Sky survey, taken by the prototype array in white light during 2006 -- 2009 and in $R$ color during 2011 -- 2012 from Chile \citep{piofthesky}. Finally and most importantly, the northern instrument of the Kilodegree Extremely Little Telescope survey (KELT-N) measured V473 Lyrae extensively during 2008 -- 2014 \citep{pepper-kelt}. The KELT-N data is the only photometry that overlaps with the \textit{MOST} run itself, providing a most important hook to determine the actual modulation phase of the star during the space-based observations. We note that data for V473 Lyrae were obtained from the KELT-N observations with simple aperture photometry instead of their differential imaging pipeline. A journal of observations is included in Table \ref{table:ground_phot_journal}. The light curves and RV data are displayed in Fig.~\ref{v473_all_lc}.

\section{Analysis of V473 L{\lowercase{yr}}}
\label{secv473}
We analyzed the \textit{MOST} and ground-based light curves with Period04 software \citep{period}. For both stars, the light curves obtained by \textit{MOST} included only a handful of pulsation cycles, resulting in low frequency resolution. Nevertheless, the high precision and quasi-continuity of the light curves provided us with advantages ground-based observations cannot offer. We present the analyis of V473 Lyr in this section, and that of U TrA in the subsequent one. Throughout the paper we designate the frequencies corresponding to the fundamental mode and the first and second overtones as $f_0$, $f_1$ and $f_2$, respectively. 

\begin{figure}
\includegraphics[width=1.0\columnwidth]{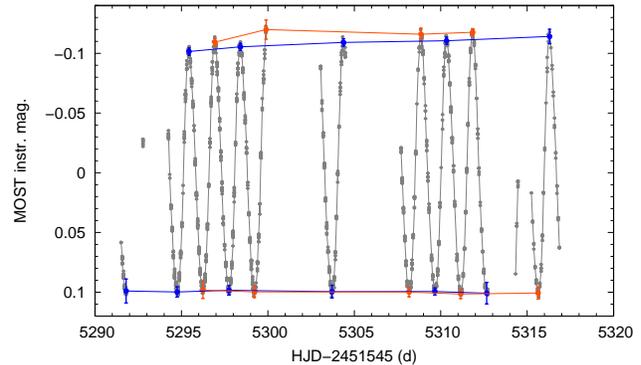}
\caption{Period doubling in V473 Lyrae. Odd and even maxima are connected with the blue and orange lines, respectively.  }
\label{v473_pd}
\end{figure}

\subsection{Period doubling in V473 Lyrae}

\begin{figure}
\includegraphics[width=1.0\columnwidth]{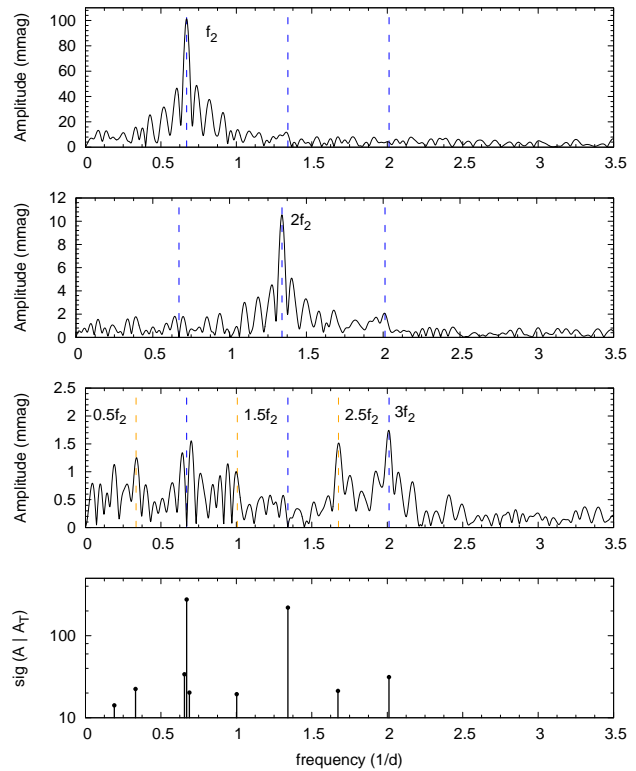}
\caption{Fourier spectra of the \textit{MOST} data of V473 Lyr. Top: full spectrum, middle: prewhitened with $f_2$, and with $f_2$ and $2f_2$. Positions of the main harmonics and subharmonics are indicated with vertical lines. Bottom: spectral significance values for all identified peaks.}
\label{v473_sp}
\end{figure}

\begin{figure*}
\includegraphics[width=1.0\textwidth]{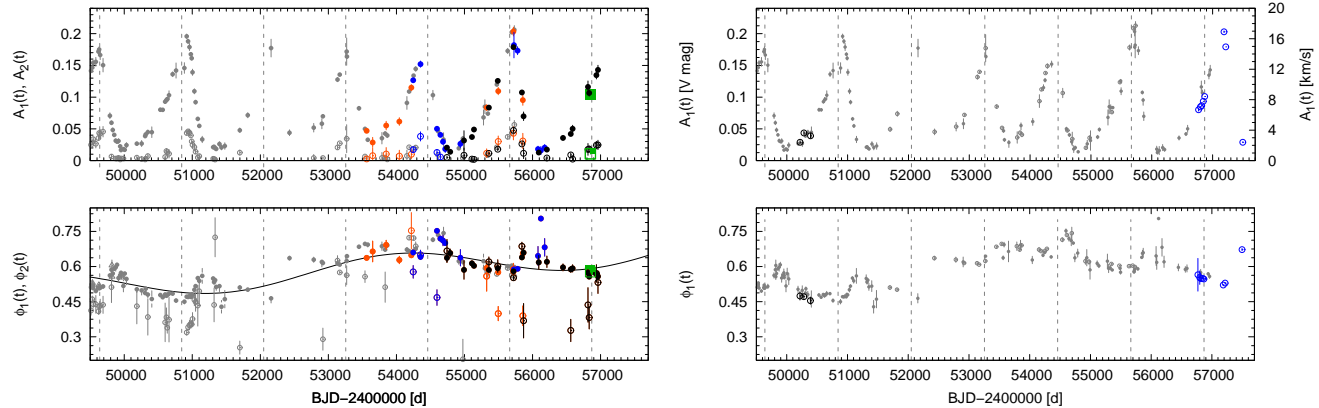}
\caption{Long-term variation of the pulsation amplitude and phase of V473 Lyrae. Top left: photometric amplitude modulation. Filled ($A_1$) and empty ($A_2$) green, orange, blue and black symbols are the \textit{MOST}, AAVSO/BSM, Pi of the Sky, and KELT-N data, respectively. Grey symbols are data from \citet{molnar2014}. Note that indices 1 and 2 correspond to the main pulsation peak ($f_2$) and its first harmonic ($2f_2$), respectively. Bottom left: the same for phase modulation. Grey dashed lines mark the amplitude maxima of the short modulation cycles, while the black line shows the long modulation cycle in the phase variation. Top right: new (blue) and some old (black) RV amplitudes overlaid on the photometric amplitudes. Bottom right: the same for RV phases. }
\label{v473_blaz}
\end{figure*}

Although the light curve of V473 Lyrae contains gaps, it samples enough pulsation extrema to confirm the presence of alternating high and low pulsation maxima. In Fig.~\ref{v473_pd}, we connect the odd and even maxima with blue and orange lines, respectively, to show that the alternation is present throughout the light curve. Alternation is observed in the maxima but not in the minima. Such repetition of two slightly different pulsation cycles is known as period doubling. The same method was used to investigate period doubling in RR Lyraes and RV Tau stars \citep{molnar2012,plachy2014}. The differences here are rather small compared to the scatter of points, therefore the method of comparing the heights of the maxima in itself gives only moderate confidence that period doubling is present. 

The Fourier spectrum of V473 Lyrae is shown in Fig.~\ref{v473_sp}. The second-overtone pulsation produces a very sinusoidal light curve and indeed we detect only the principal pulsation frequency and its two harmonics. The presence of period doubling leads to the appearance of half-integer frequencies $((2n+1)/2\,f_2)$ in the spectrum. In this case we are able to detect only a single additional peak at $2.5\,f_2$ with high ($>4$) signal--to--noise ratio (SNR). The 0.5 and $1.5\,f_2$ peaks can also be identified but their SNR may fall below 4, the canonically accepted value in the literature, depending on the exact calculation of the noise level. After the subtraction of the frequencies listed in Table \ref{fourtbl-v473}, the rms scatter of the residual was found to be 3.1 mmag. 

However, the low SNRs of the frequency components can be misleading in this case. Estimating the SNR, especially for low-frequency signals, can be unreliable as it depends on the level of prewhitening applied to the data. Moreover, here the pulsation amplitude and phase of the star were changing slightly over the \textit{MOST} observations due to the modulation, and thus the residual spectrum contains small side peaks that account for these changes and may be difficult to properly prewhiten. 

Therefore instead of relying only on the SNR estimate alone, we also calculated the spectral significance values with SigSpec \citep{sigspec}. The spectral significance is just the logarithm of the inverse of the false-alarm probability of the frequency component. The analysis revealed three half-integer components (0.5, 1.5, and 2.5\,$f_2$) with significance values between 19 and 22, clearly confirming the presence of period doubling in the star. Side peaks around the $f_2$ peak are also detected with high significance, as expected from the modulation-induced variations. One more peak can be identified with high significance level (sig = 14.1) at $f' = 0.19105$ d$^{-1}$, but it is likely spurious, as it is not detected if the data are binned per orbit. 

To further confirm the presence of period doubling, we also checked the frequency content of the data obtained for the guide star used by \textit{MOST}, HD 337939 ($V=9.55$ mag). A single low-amplitude peak can be identified at 1.00 d$^{-1}$ that is likely caused by the daily variation in the amount of stray light. Although this frequency coincides with the $1.5\,f_2$ subharmonic peak we identified in V473 Lyr, and has a similar amplitude, it cannot explain the presence of the other two subharmonics, and its phase is also significantly different from that of the $1.5\,f_2$ peak in V473 Lyr. We can conclude that stray light may add uncertainty to the amplitude and phase of the $1.5\,f_2$ peak but otherwise has no adverse effects on the detection of period doubling. 

Alternating minima or maxima were first identified in RV Tau stars \citep{seares1908,vanderbilt1916} and were discovered in other Type II Cepheids and in RR Lyrae stars much later \citep{templeton2007,szabo2010,smolec2012,plachy2016}. With this detection, classical Cepheids, at least those pulsating in the second overtone, join the group of other classical pulsators where period doubling has been detected.

\subsection{Ground-based photometry}

The data we acquired from the three different ground-based surveys are shown in Fig.~\ref{v473_all_lc}. As seen in this figure, the coverage and quality of the measurements are very different. The Pi of the Sky data in particular suffer from large photometric uncertainties. The pulsation amplitudes are also different, partly because of the use (or lack) of different filters, and contamination from nearby stars. However, we were able to determine the temporal variation of the $A_1$ and $\phi_1$, and in some cases, the $A_2$ and $\phi_2$ Fourier parameters from all data sets. (Here the indices 1 and 2 refer to frequency components $f_2$ and $2f_2$.) We then scaled the amplitudes to the values presented by \citet{molnar2014}. 

The amplitude and phase modulation curves are displayed in Fig.~\ref{v473_blaz}. The new observations confirm that a strong amplitude maximum occurred around JD 2455700, with $A_1$ exceeding 0.2 mag in the $V$ band. Comparison of the very end of the data set with the \textit{MOST} data point suggests that the pulsation amplitude was still increasing during the space-based observations, although they were obtained very close to the calculated time of the amplitude maximum. The shape of the phase modulation ($\phi_1(t)$) curve changed from an upward to a downward slope sometime around JD 2454500, following the extrapolated sine fit of the secondary modulation determined by \citet{molnar2014}. This further supports that a long-period secondary modulation is at work in the star. 

\subsection{Radial velocity measurements}

RV variations of V473 Lyrae were monitored for an extended period of time by \citet{burki2006}, but measurements ceased after 1997 (JD 2450568). \citet{molnar2014} showed that the changes in the amplitudes and phases of earlier RV data follows the modulation observed in the photometric amplitudes and phases closely. Comparison of our new photometric and RV data sets shows that this is still the case. The observations indicate that the RV variations follow both the short and long modulations in accordance with the photometry, confirming that both modulations are intrinsic to the pulsation of the star. We compared the pulsation-averaged, or $\gamma$-velocities of the available data sets to search for any signs of binarity. The scatter of points does not exceed 0.3 km s$^{-1}$ for our campaigns. Older data sets show considerably larger deviations, but even those are in the range of 1.3 km$^{-1}$. Assuming that either of the phase modulations are caused by light-time effect, much larger projected velocities would be required to account for the observed differences. We conclude that V473 Lyrae has no binary companions, at least with orbital periods shorter than several decades. 

\subsection{Phase lag and physical parameters}
\begin{figure}
\includegraphics[width=0.97\columnwidth]{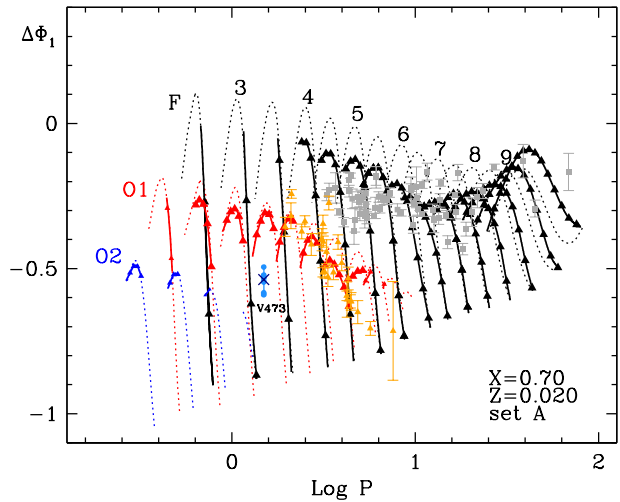}
\caption{Phase lag of V473 Lyrae (light blue dots are our analysis of the older data sets, blue cross is for the \textit{MOST} run), compared to the model calculations of \citet*{szabo2007}. Grey and orange symbols are the fundamental-mode and first-overtone Cepheids collected by \citet{ogloza2000}.  The star is clearly separated from the first-overtone ones and follows the second-overtone model family.   }
\label{phase_lag}
\end{figure}

The phase lag ($\Delta \Phi_1$) is the difference in phase between the luminosity and velocity variations. For their model calculations, \citet{szabo2007} defined it as the difference between the phase of the light curve and the star-centric velocity that is the opposite of the measured RV. At the time, there were no observations of second overtones with which they could compare their models. Although extensive photometric and spectroscopic data have been collected separately about V473~Lyr during the past decades, the phase lag of the star was not calculated before our work. Since the star's pulsations are modulated, we selected various modulation phases with coincident (or nearly coincident) photometric and spectroscopic observations. The phase lag indeed shows some variation between $\Delta \Phi_1 = -0.49\dots-0.59$, decreasing from low to high pulsation amplitudes. We found the phase lag to be $\Delta \Phi_1 = -0.538 \pm 0.012$ during the \textit{MOST} run that was coincident with our first spectroscopic campaign. These values, with the pulsation period of ${\rm log} P({\rm d})  = 0.1734$ d, put the star well below the first overtone sequence, as Figure \ref{phase_lag} illustrates. Although V473 Lyrae has a longer pulsation period than the second-overtone regime presented by \citet*{szabo2007}, the extent of the nonlinear regime depends on the convective parameters of the models, and different parameters could extend it to longer periods. Our result supports the findings of \citet{molnar2014} that the star pulsates in the second overtone.

Finally, we derived the fundamental stellar parameters of the star from the spectra according to the recipe described by \citet{moor}. We summed 20 individual spectra taken at KUO on 16 July 2015 (HJD 2457219) and compared this master spectrum with templates taken from the spectral library of \citet{munari2005}.  We used an iterative method to estimate the stellar parameters. After transforming the measured spectrum to the laboratory system, we compared it to a grid of Munari synthetic spectra in the 4400--5500 \AA{} range, excluding the H$\beta$ region. The grid was compiled by varying effective temperature, surface gravity, and projected rotational velocity of the model spectra, while the metallicity was fixed at [Fe/H] = 0.0. After finding the global minimum we used the derived $T_{\rm eff}$ and ${\rm log}\, g$ parameters to determine M/H metallicity and $v\sin\, i$. This iterative approach is stable against the parameter degeneracies, especially between $T_{\rm eff}$ and M/H. By linear interpolation around the minimum, we derived the final parameter set as: $T_{\rm eff} = 6175 \pm150$ K, ${\rm log}\, g = 2.7 \pm 0.2$, $v {\rm sin}\, i = 17 \pm 5$ km~s$^{-1}$. These values agree well with those derived by \citet{Andrievsky}.

\section{Analysis of U T{\lowercase{r}}A}
\label{secutra}

\subsection{Possible detection of $\boldmath{f_X}$}
\begin{figure}
\includegraphics[width=1.0\columnwidth]{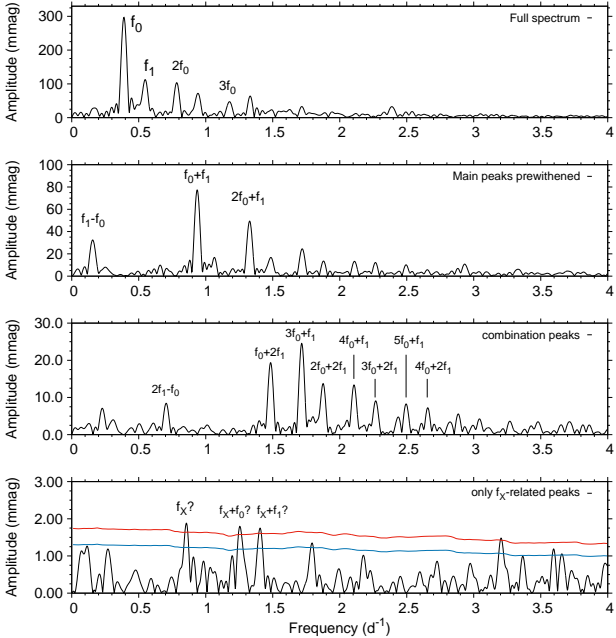}
\caption{Fourier spectra of the \textit{MOST} data of U TrA. Red and blue lines indicate SN ratios 4 and 3. }
\label{utra_sp}
\end{figure}

The \textit{MOST} light curve of U TrA is very well sampled and shows the characteristic beating pattern of double-mode pulsation. Although the frequency resolution is limited by the short length of the data, the Fourier spectrum revealed numerous combination peaks between the fundamental mode and the first overtone (Fig.~\ref{utra_sp}). The rms scatter of the residual light curve after prewhitening with the frequencies listed in Table~\ref{fourtbl-utra} is 5.7 mmag.

Observations of the OGLE survey revealed that many first-overtone Cepheids both in the Galactic Bulge and in the Magellanic Clouds show one or more low-amplitude additional modes in the period range range of $P/P_1 \sim 0.61$~--~$0.64$ \citep{moskalik2009,soszynski2008,soszynski2010}. The same modes were also identified in a few beat Cepheids \citep{soszynski2015}. First-overtone and double-mode RR Lyrae stars also show very similar modes. Notations used for these modes in the literature are $f_{0.61}$ or, simply $f_X$.

We tentatively identify peaks in the frequency spectrum of U TrA that may correspond to the $f_X$ mode at 0.8615 d$^{-1}$, with a frequency ratio of $f_1/f_ X = 0.636$, as shown in the lowest panel of Fig.~\ref{utra_sp}. Two other peaks can be identified as the $f_X+f_0$ and $f_X+f_1$ combinations, respectively. All peaks have SNRs larger than 4 (red line). Unlike in the case of V473 Lyr, the mode amplitudes appear to be constant, hence the SNR can be calculated more accurately for U TrA.

However, given the short data set and hence the low frequency resolution, the three peaks we associated with the $f_X$ mode can be fitted as combination peaks too. The position of the suspected $f_X$ peak and the nearest combination, $5f_0-2f_1$, differ only by 0.010 d$^{-1}$, much less than the actual $\Delta f = 1/\Delta T = 0.037$ d$^{-1}$ frequency resolution of the \textit{MOST} light curve. This also means that the $f_X+f_1$ peak overlaps with the much stronger $5f_0+f_1$ peak, but we note that a common frequency fit does assign significant amplitude to both peaks. 

Overall, although it is plausible that U TrA, as many Cepheid stars pulsating in the first overtone, contains the $f_X$ mode, the overlaps with the combination peaks and the limited frequency resolution prevents us from an unambiguous identification. A decisive study would require much longer observations to disentangle the nearby frequency components. Unfortunately, data on U TrA from the next space-based photometric mission, \textit{TESS}, will most likely also be limited to 27 days, i.e., the same length as the \textit{MOST} data. Other beat Cepheids that are closer to the ecliptic will be observed during the K2 mission of the \textit{Kepler} space telescope that provides us with longer observations and higher photometric accuracy, making it possible to search for the $f_X$ mode in other stars too \citep{molnar-k2}.

\subsection{Radial velocity measurements}
\label{sec-utra-rv}

In the case of U TrA, time-resolved RV measurements are scarce: data presented by \citet{stibbs1955}, \citet{laustsen1957}, and \citet{usenko2014} are limited to a handful of points. Only \citet{stobie1979} provided observations with good phase coverage. Unfortunately, our coverage was sparse as well. Nevertheless, we compared these measurements and fitted them with the pulsation periods determined from the photometric data (see also Section \ref{energy-oc}) to search for any signs of $\gamma$-velocity variations that would indicate orbital motion caused by a binary companion. We did not detect any systematic variations larger than $\pm 4$ km\,s$^{-1}$, the scatter of the residual RV data. 

\begin{figure}
\includegraphics[width=1.0\columnwidth]{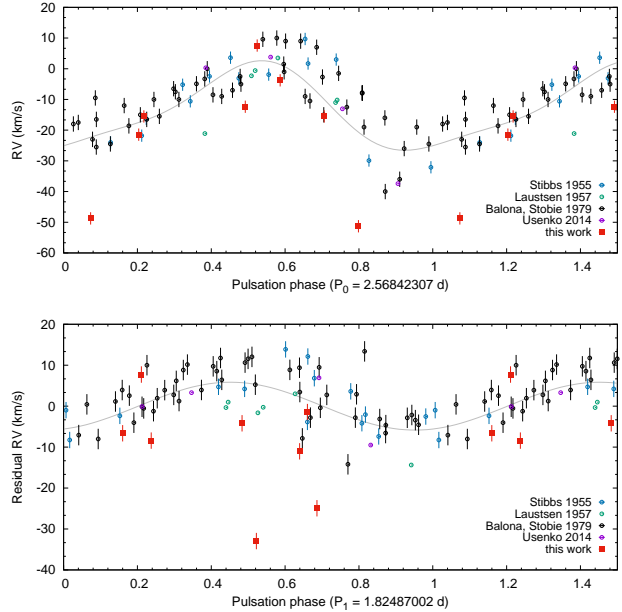}
\caption{Archival (grey, blue, purple circles), and newly obtained SASER RV data (red sqares) of U TrA, phased with the two pulsation periods. Top: original data set, with a two-component fit of the fundamental mode, the full amplitude of the curve is $A_{\rm FM} = 29.1$ km\,s$^{-1}$. Bottom: residual with a fit to the first overtone, $A_{\rm 1O} = 11.7$ km\,s$^{-1}$. The two outlier points around --50 km\,s$^{-1}$ were not used when fitting the amplitudes.}
\label{utra_rv_plot}
\end{figure}

\begin{figure}
\includegraphics[width=1.0\columnwidth]{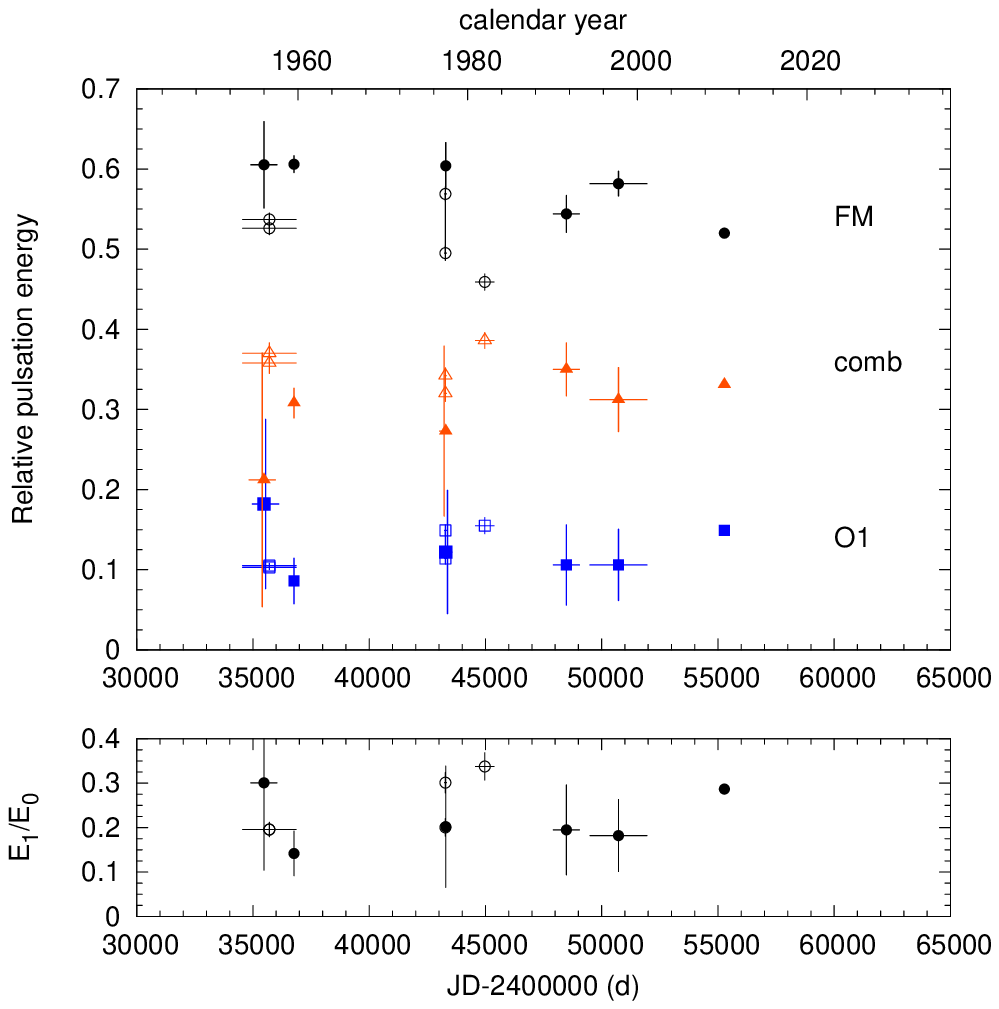}
\caption{Top: relative pulsation energy levels in U TrA (sum of the energy terms corresponding to each group divided by the total sum of the energy terms). Empty symbols are the values calculated \citet{faulknershobbrook79} and \citet{faulknershobbrook83}; data that were analysed in both studies are connected with vertical lines. Filled symbols: our analysis of the following data sets: \citet{jansen1962}, \citet{mitchell1964}, \citet{stobie1979}, Hipparcos  \citep{esa}, \citet{berdnikov2008} and \textit{MOST}. Bottom: $E_1/E_0$ energy relation of the first overtone and the fundamental mode.  }
\label{utra_energy}
\end{figure}

\subsection{Search for energy exchange and period variations}
\label{energy-oc}

\citet{faulknershobbrook79} raised the possibility of pulsation energy exchange between the two pulsation modes via mode coupling in U TrA. To see if any change has occurred in the almost four decades since, we analysed Hipparcos photometry, observations by \citet{mitchell1964,berdnikov2008}, and the \textit{MOST} light curve. We also re-analysed photometry by \citet{jansen1962} and \citet{stobie1979}. Following the notation of \citet{faulknershobbrook79}, we define the energy terms as $\sum (f_{ij}A_{ij})^2$, where $i$ and $j$ refer to the pulsation modes. Indices $i=j=0$ correspond to peaks belonging to the fundamental mode, $i=j=1$ to the first overtone, and $i\neq j$ to the combination (or cross-coupling) terms in the Fourier solution. 

The relative energy levels we calculate show considerable scatter from one data set to another, as can be seen in Fig.~\ref{utra_energy}. Despite the scatter, we see no evidence of energy exchange between the two pulsation modes on a decadal time scale. The energy level of the first overtone remains essentially flat over the time spanned by the data, although some variation can be seen between the fundamental mode and the combination peaks. Considering the scatter of points, we conclude that the relative pulsation energy levels of the fundamental mode and the first overtone did not experience changes exceeding $\pm 7$ and $\pm 5$ per cent, respectively. The derived ratio of the energy levels of the two modes (lower panel of Fig.~\ref{utra_energy}) is more uncertain, the scatter of points reach $\pm 13$ per cent.

\begin{figure}
\includegraphics[width=1.0\columnwidth]{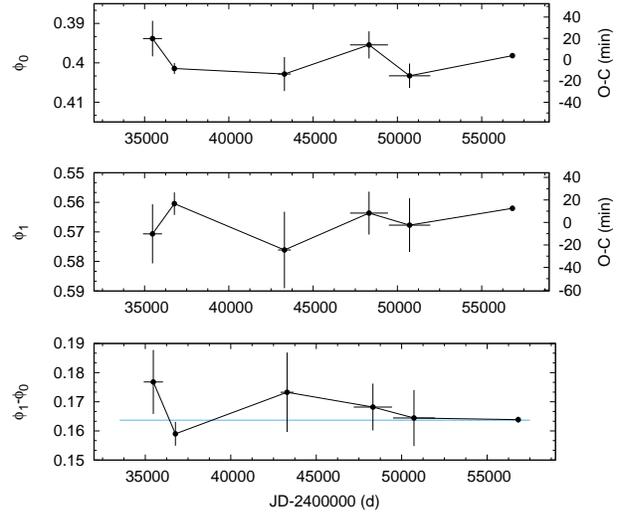}
\caption{Pulsation phase variations in U TrA. Top: fundamental mode, middle: first overtone, bottom: phase difference between the two modes. The horizontal line marks the average difference of $\phi_1-\phi_0=0.1637$. Units are in rad/$2\pi$. The reference periods we used are: $P_0 = 2.56842307$~d and $P_1 = 1.82487002$~d. }
\label{utra_phasevar}
\end{figure}

We also investigated the stability of the pulsation periods in U TrA. Classical O--C methods, based on the timings of the extrema of the cycles, or on the moments of the median brightness on the ascending branch of the light curve, are less effective for beat Cepheids, so period changes in these stars have been rarely explored. Here, we fit the available data sets with the same frequency components and estimate changes in pulsation phases of the fundamental mode and first overtone ($f_0$ and $f_1$) simultaneously. We also determine the pulsation periods to high accuracy by minimising the linear slopes of the two O--C curves: $P_0 = 2.56842307$~d and $P_1 = 1.82487002$~d. 

Fig.~\ref{utra_phasevar} shows that neither mode in U TrA undergoes clear variations in time, nor does the $\phi_1-\phi_0$ difference in phase between them. This suggests a lack of nearby companion stars, in agreement with the RV data, and very stable pulsation periods, in agreement with the findings of \citet{pardoporetti}. There is some scatter in the phase of the fundamental mode from one data set to another, but that is likely due to uncertainties arising from  their sparse samplings. We note that \textit{HST} direct imaging revealed two candidate companions to U TrA, but the general conclusion of the survey is that stars with separations as wide as these are not physical companions. In any case, if they were companions, they would have orbital periods in the range of $10^5$--$10^6$ yr around the Cepheid, so their effects would be undetectable in our timing data \citep{evans2016}.

\section{Discussion and Summary}
\label{secsum}
\subsection{Additional modes in Cepheids}

The detection of low-amplitude additional modes in Cepheid and RR Lyrae stars, often below mmag level, was a major advance in the field of classical pulsating stars in the last decade (see, e.g., \citealt{gruberbauer2007,soszynski2008,benko2010,molnar2016}). While some modes have been explained \citep{kollathpd,kollath2016}, the mechanism for others, like the $f_X$ mode(s), between $P_X/P_0 \sim 0.60-0.64$, are still disputed. The most promising hypothesis was presented by \citet{dziem2016} who attributed the signal at $f_X$ to the first harmonic of three different non-radial modes at frequencies of $f_X/2$ with orders of $\ell=7-9$. These modes experience strong cancellation at the pulsation frequency, making the $f_X = 2\,f_\ell$ component the strongest one observationally. If the peak we found in U TrA is indeed such a mode, it would belong to the central, $\ell=8$ group. 

We did not detect additional modes in V473 Lyrae directly, but based on prior experience, the presence of period doubling strongly suggests that at least one more mode is excited in the star, and is in resonance with the main mode. Our hypothesis is based on the assumption that even though the evolutionary states of RR Lyrae and various Cepheid stars are very different, the structure of their envelopes and thus their pulsations are remarkably similar. In fact, radial pulsations in RR Lyrae, classical, and type II Cepheids can be modeled with minimal changes to the same hydrodynamic codes \citep{kollath2002,smolecmoskalik}. 

In the case of RR Lyrae stars, the origin of period doubling was unambiguously traced back to the $P_0:P_9 =9:2$ resonance between the fundamental mode and the 9th overtone, which is a strange mode \citep{kollathpd}. For Type II Cepheids, nonlinear hydrodynamical models point towards a 3:2 resonance with the first overtone and a 5:2 resonance with the second overtone for the BL Her and W Vir subtypes, respectively, and low-amplitude modulation was also identified \citep{smolecmoskalik2012,smolec2016}. In the case of second-overtone classical Cepheids, only linear model calculations have been carried out so far, suggesting that resonances can occur in the vicinity of the physical parameters of V473 Lyrae \citep{molnar2014}. The best candidates are the 4th and 8th overtones with possible 3:2 or 5:2 resonances, but a nonlinear model survey is needed to pin down the exact mode that may cause the period doubling. Nevertheless, the simultaneous presence of the modulation and period doubling shows a striking similarity to what we see both in RR Lyrae stars and type II Cepheid models, therefore we consider it to be plausible that both phenomena may originate from mode resonances in RR Lyrae and Cepheid stars alike \citep{bk11}.  

\subsection{Summary}

We analysed the light curves of the Cepheids V473 Lyr and U TrA obtained by the \textit{MOST} space telescope, supported by ground-based photometry and spectroscopy. The \textit{MOST} observations provided us with one month of each star, with nearly continuous coverage of U TrA and 57 per cent of V473 Lyrae (but nearly continuous coverage outside the long gaps). 

Our space-based photometry produced new insights into both stars:

\begin{itemize}
\item We see period doubling (alternating higher and lower maxima) in V473 Lyrae, the first detection of this phenomenon in a classical Cepheid star. The combination of strong amplitude and phase modulation (a clear phenomenological analog to the Blazhko effect seen in RR Lyrae stars) and period doubling suggests that mode resonances could be the common explanation for modulation in Cepheids and RR Lyrae stars \citep{bk11}.

\item In U TrA, we tentatively identify one frequency component as an $f_X$ mode, but because of the limited frequency resolution of the data, we cannot exclude the possibility that it is instead a combination peak. 
\end{itemize}

\noindent The combination of the precise but relatively short {\textit MOST} light curves with longer but much sparser and lower-precision ground-based photometric data enabled us to reach the following conclusions about the changes in the target stars over decadal time scales:

\begin{itemize}
\item Data from the AAVSO BSM, Pi of the Sky and KELT-N surveys fill the large gap between the last observations of V473 Lyrae in \citet{molnar2014} and the \textit{MOST} run. Together, they show that V473 Lyrae did not reach maximum pulsation amplitude by the time of the \textit{MOST} run, due to the phase shift caused by the secondary modulation. 

\item Observations of U TrA, one of the first beat or double-mode Cepheids discovered, extend back to the 1950s. Analysis of all available data points to no energy exchange (larger than 5 per cent) between the fundamental mode and the first overtone, disproving the suggestion made by \citet{faulknershobbrook79}. We also found no significant changes in the phase of either pulsation period. 
\end{itemize}

\noindent Spectra of both stars (from a multi-site campaign for V473 Lyrae, and a handful of observations for U TrA) added the following results: 

\begin{itemize}
\item We measured the phase lag between the luminosity and RV variations in V473 Lyrae. This is the first time for a second overtone Cepheid. Our results agree very well with the theoretical predictions of \citet*{szabo2007}. We refined the following fundamental physical parameters of V473 Lyrae: $T_{\rm eff} = 6175 \pm150$ K, ${\rm log}\, g = 2.7 \pm 0.2$, $v {\rm sin}\, i = 17 \pm 5$ km~s$^{-1}$.

\item In our new RV data for three different modulation phases of V473 Lyr (during amplitude increase near maximum amplitude and on the descending branch), combined with archival spectra, we detect no $\gamma$-velocity changes larger than 0.3 and 1.3 km\,s$^{-1}$ that could indicate binarity. 

\item Our few new RV measurements for U TrA, combined with archival spectra, do not show changes in the $\gamma$-velocity of the star beyond $\pm 4$ km\,s$^{-1}$. 
\end{itemize}

The detection of period doubling in a modulated Cepheid raises the possibility of a unified mode resonance model for the Blazhko effect in RR Lyrae and Cepheid stars. V473 Lyrae will be observed by the \textit{TESS} space telescope to further investigate this phenomenon. \textit{TESS} will be able to observe the star for about the same length of time as \textit{MOST} (about 1 month), although V473 Lyr may fall into the overlapping part of two neighboring sectors of the \textit{TESS} observing scheme, extending the possible coverage. 

Other bright Cepheids have been and will be observed by other space photometry missions (\textit{Kepler} and \textit{BRITE} \citep{brite}) and new observations are possible with \textit{MOST} to look for such effects. \textit{TESS} will observe several Cepheids, including stars like Polaris and the archetype $\delta$ Cep. But given  long pulsation periods for some Cepheids, extending to weeks or months, these measurements are sometimes just snapshots giving tantalizing glimpses of the underlying dynamics (see, e.g., \citealt{plachy2016}). 

An important takeaway from our study is the level of benefit of combining space-based photometry with extended ground-based photometry and spectroscopy. While current ground-based photometric surveys designed to search for exoplanets and asteroids (such as KELT) provide valuable data, other projects aimed at simultaneous all-sky coverage and higher duty cycles have even greater potential to investigate Cepheids. New survey machines like MASCARA, Evryscope \citep{mascara,evryscope} and especially the multicolour Fly's Eye project \citep{flyseye} will be natural extensions of the \textit{TESS} mission. We believe combining the results of all these tools is the best way forward to further our understanding of the physical properties of Cepheids and their pulsations.

\section*{Acknowledgments}
We thank Matthew Templeton for his assistance with the AAVSO observations of V473 Lyr. This project has been supported by the Lend\"ulet LP2012-31 and LP2014-17 Programs of the Hungarian Academy of Sciences, and by the NKFIH K-115709, K-119517, PD-116175, and PD-121203 grants of the Hungarian National Research, Development and Innovation Office. The research leading to these results has received funding from the ESA PECS Contract No.\ 4000110889/14/NL/NDe,  and from the European Community's Seventh Framework Programme (FP7/2007-2013) under grant agreement no.\ 312844 (SPACEINN). L.M.\ and \'A.S. were supported by the J\'anos Bolyai Research Scholarship of the Hungarian Academy of Sciences.  AD has been supported by the Postdoctoral Fellowship Programme of the Hungarian Academy of Sciences and by the Tempus K\"ozalap\'itv\'any and the M\'AE\"O. AD, BCs, JK and GyMSz would like to thank the City of Szombathely for support under Agreement No. 67.177-21/2016. MS acknowledges the support of the postdoctoral fellowship programme of the Hungarian Academy of Sciences at the Konkoly Observatory, MTA CSFK as a host institution. NDR is grateful for postdoctoral support by the University of Toledo and by the Helen Luedtke Brooks Endowed Professorship. AFJM and JMM are grateful for financial aid from NSERC (Canada) and FQRNT (Qu\'ebec).This publication is based (in part) on spectroscopic data obtained through the collaborative Southern Astro Spectroscopy Email Ring (SASER) group. This research has made use of the SIMBAD database, operated at CDS, Strasbourg, France, and NASA's Astrophysics Data System.


\appendix
 
\section{Data tables}

This appendix contains various data tables used in the paper. Radial velocity measurements are presented in Tables \ref{v473-vr-tbl} and \ref{utra-sp-log-tbl}. The observations of the \textit{MOST} space telescope are listed in Table \ref{v473-most-tbl} and \ref{utra-most-tbl} for V473 Lyr and U TrA, respectively. Tables \ref{fourtbl-v473} and \ref{fourtbl-utra} contains the Fourier coefficients we determined for V473 Lyr and U TrA. 

\begin{table*}   
\begin{center}
 \caption{Journal of spectroscopic observations of U TrA by the members of SASER and from ICRAR (Luckas). Top: spectra, with successful RV determination, time is corrected to BJD. Values in parentheses are offset from the literature data and likely erroneous. Bottom: spectra where we could not determine reliable RV values. Time is in JD.}  
\label{utra-sp-log-tbl}  
\begin{tabular}{cccccc}  
\hline   
Time & RV & Observer & Detector & Wavelength range & Resolution \\
 (BJD/JD) & (km s$^{-1}$) & ~            &  ~            & (\AA) & $\Delta \lambda / \lambda$ \\
\hline  
2456817.10503 & --12.37 & Heathcote &  & 4570--4792 & 5343 \\
2456818.96822 & --15.53 & Heathcote &  & 4571--4791 & 6754 \\
2456851.04387 & --15.53 & Heathcote &  & 4573--4791 & 4011 \\
2456860.02710 & --21.47 & Heathcote &  & 4577--4792 & 6125 \\
2457173.04020 & (--48.7) & Luckas &  & 5641--5774 & 9408 \\
2457180.04022 & (--51.3) & Luckas &  & 6065--6189 & 12884 \\
2457200.04302 & --3.76 & Bohlsen &  & 6080--6606 & 7606 \\
2457205.02001 & +7.56 & Bohlsen &  & 3799--7198 & 1464 \\
\hline
2457174.991 & -- & Bohlsen &  & 3799--7199 & 1159 \\
2457194.578 & -- & Cacella &  & 5785--6298 & 4289 \\
2457194.585 & -- & Cacella &  & 5785--6298 & 4289 \\
2457194.592 & -- & Cacella &  & 5785--6298 & 4289 \\
2457194.600 & -- & Cacella &  & 5785--6298 & 4289 \\
2457224.998 & -- & Bohlsen &  & 3799--7199 & 1504 \\
\hline
\end{tabular} 
\end{center}  
\end{table*}

\begin{table}   
\begin{center}
 \caption{Radial velocity data of V473 Lyr. This is a sample only, the entire table is accessible in the online version of the paper. }  
\label{v473-vr-tbl}  
\begin{tabular}{cccc}  
\hline   
 BJD--2450000 & RV & Uncertainty & Observatory  \\
  (d)      &  (km s$^{-1}$)    &   (km s$^{-1}$)  & ~    \\
\hline  
6760.500400 & -12.80 & 0.12 & GAO \\
6760.507449 & -12.86 & 0.14 & GAO \\
6760.514498 & -12.73 & 0.12 & GAO \\
6760.521559 & -12.46 & 0.13 & GAO \\
6760.528608 & -12.24 & 0.14 & GAO \\
\multicolumn{3}{l}{\dots} \\
\hline
\end{tabular} 
\end{center}  
\end{table}

\begin{table}   
\begin{center}
 \caption{Photometric data of V473 Lyr obtained by the \textit{MOST} space telescope. This is a sample only; the entire table is accessible in the online version of the paper. Units are in the instrumental magnitudes of \textit{MOST}. }  
\label{v473-most-tbl}  
\begin{tabular}{cccc}  
\hline   
 HJD--2451545 & Brightness  & Uncertainty  \\
  (d)      &  (\textit{MOST} mag)    &    (\textit{MOST} mag)     \\
\hline  
5291.507433 & 12.19546 & 0.00043 \\
5291.508872 & 12.19586 & 0.00043 \\
5291.573474 & 12.20719 & 0.00043 \\
5291.574194 & 12.20830 & 0.00043 \\
5291.574913 & 12.20876 & 0.00043 \\
\multicolumn{3}{l}{\dots} \\
\hline
\end{tabular} 
\end{center}  
\end{table}

\begin{table}
\begin{center}
 \caption{Photometric data of U TrA obtained by the \textit{MOST} space telescope. This is a sample only; the entire table is accessible in the online version of the paper. Units are in the instrumental magnitudes of \textit{MOST}. }  
\label{utra-most-tbl}  
\begin{tabular}{cccc}  
\hline   
 HJD--2451545 & Brightness  & Uncertainty  \\
  (d)      &  (\textit{MOST} mag)    &    (\textit{MOST} mag)     \\
\hline  
5261.701539 & 13.38936 & 0.00054 \\
5261.702972 & 13.39279 & 0.00054 \\
5261.707270 & 13.38989 & 0.00054 \\
5261.708703 & 13.38948 & 0.00054 \\
5261.710136 & 13.38934 & 0.00053 \\
\multicolumn{3}{l}{\dots} \\
\hline
\end{tabular} 
\end{center}  
\end{table}

\begin{table}   
\begin{center}
 \caption{Results of the frequency analysis of V473 Lyr.}  
\label{fourtbl-v473}  
\begin{tabular}{cccc}  
\hline   
 ID & ~Frequency~ & ~Amplitude~ & Phase  \\
  ~ &     (d$^{-1}$)      &  (mag)    &    ~(rad/2$\pi$)~      \\
\hline    
 $f_2$  &	0.67071(29)	&	0.10464(13)	&	0.0482(20)	\\
 $2f_2$  &	1.3416(25)	&	0.01048(14)	&	0.3372(23)	\\
 $3f_2$ &	2.0116(30)	&	0.00157(40)	&	0.560(12)	\\
$0.5\,f_2$  &	0.3359(30)	&	0.0016(5)	&	0.600(12)	\\
$1.5\,f_2$  &	1.0000(19)	&	0.00137(15)	&	0.768(12) 	\\
$2.5\,f_2$  &	1.6754(25)	&	0.00161(22)	&	0.155(16)	\\
\hline
\end{tabular} 
\end{center}  
\end{table}

\begin{table}   
\begin{center}
 \caption{Results of the frequency analysis of U TrA.}  
\label{fourtbl-utra}  
\begin{tabular}{cccc}  
\hline   
 ID & ~Frequency~ & ~Amplitude~ & Phase  \\
  ~ &     (d$^{-1}$)      &  (mag)    &    ~(rad/2$\pi$)~      \\
  \hline 
$f_0$	&	0.389377(8)	&	0.30050(15)	&	0.44966(6)	\\
$2f_0$	&	0.77872(2)	&	0.09939(14)	&	0.49871(18)	\\
$3f_0$	&	1.16826(6)	&	0.03864(14)	&	0.4693(5)	\\
$4f_0$	&	1.5581(2)	&	0.01512(14)	&	0.8976(12)	\\
$5f_0$	&	1.9461(4)	&	0.00627(15)	&	0.876(3)	\\
$6f_0$	&	2.3355(9)	&	0.00272(14)	&	0.975(7)	\\
$f_1$	&	0.54797(2)	&	0.11392(13)	&	0.57758(16)	\\
$2f_1$	&	1.09606(16)	&	0.01471(13)	&	0.9631(12)	\\
$3f_1$	&	1.6511(9)	&	0.0025(9)	&	0.736(7)	\\
$f_0+f_1$	&	0.93721(3)	&	0.07655(14)	&	0.1290(2)	\\
$2f_0+f_1$	&	1.32686(5)	&	0.04889(14)	&	0.5661(4)	\\
$3f_0+f_1$	&	1.71587(9)	&	0.02463(15)	&	0.4039(7)	\\
$4f_0+f_1$	&	2.10569(18)	&	0.01305(14)	&	0.9232(14)	\\
$5f_0+f_1$	&	2.4944(3)	&	0.00696(14)	&	0.367(3)	\\
$6f_0+f_1$	&	2.8820(6)	&	0.00364(14)	&	0.678(5)	\\
$7f_0+f_1$	&	3.2639(15)	&	0.00159(14)	&	0.630(11)	\\
$f_0+2f_1$	&	1.48570(12)	&	0.01891(15)	&	0.4091(10)	\\
$2f_0+2f_1$	&	1.87547(18)	&	0.01328(14)	&	0.2408(14)	\\
$3f_0+2f_1$	&	2.2640(2)	&	0.00937(15)	&	0.351(2)	\\
$4f_0+2f_1$	&	2.65331(4)	&	0.00633(13)	&	0.762(3)	\\
$5f_0+2f_1$	&	3.0394(7)	&	0.00324(15)	&	0.886(6)	\\
$6f_0+2f_1$	&	3.4318(14)	&	0.00163(14)	&	0.817(11)	\\
$7f_0+2f_1$	&	3.8180(16)	&	0.00147(13)	&	0.566(12)	\\
$8f_0+2f_1$	&	4.2101(19)	&	0.00118(15)	&	0.702(15)	\\
$f_0+3f_1$	&	2.0293(7)	&	0.00354(13)	&	0.539(5)	\\
$2f_0+3f_1$	&	2.4247(6)	&	0.00402(13)	&	0.604(5)	\\
$3f_0+3f_1$	&	2.8141(9)	&	0.00277(15)	&	0.414(7)	\\
$f_1-f_0$	&	0.15827(7)	&	0.03331(15)	&	0.6674(5)	\\
$2f_0-f_1$	&	0.2304(3)	&	0.00691(13)	&	0.627(3)	\\
$3f_0-f_1$	&	0.6158(10)	&	0.00241(14)	&	0.515(8)	\\
$4f_0-f_1$	&	1.0040(8)	&	0.00314(14)	&	0.408(6)	\\
$5f_0-f_1$	&	1.4024(12)	&	0.0020(3)	&	0.968(9)	\\
$2f_1-f_0$	&	0.7070(3)	&	0.00755(14)	&	0.453(2)	\\
$2f_1-2f_0$	&	0.3212(7)	&	0.00320(14)	&	0.398(6)	\\
$f_X$?	&	0.8551(13)	&	0.00185(13)	&	0.283(10)	\\
$f_X+f_1$?	&	1.2632(9)	&	0.0027(3)	&	0.568(7)	\\
$f_X+f_0$?	&	1.27602(15)	&	0.00161(16)	&	0.417(11)	\\
\hline
\end{tabular} 
\end{center}  
\end{table}

\pagebreak

\subsection*{Affiliations:}
$^{1}$Konkoly Observatory, MTA CSFK, Konkoly Thege Mikl\'os \'ut 15-17, H-1121 Budapest, Hungary\\
$^2$ELTE Gothard Astrophysical Observatory, H-9704 Szombathely, Szent Imre herceg \'ut 112, Hungary\\
$^3$Department of Physics and Astronomy, University of British Columbia, 6224 Agricultural Road, Vancouver, BC V6T 1Z1, Canada\\
$^{4}$Department of Mathematics, Physics \& Geology, Cape Breton University, 1250 Grand Lake Road, Sydney, Nova Scotia, Canada, B1P 6L2\\
$^{5}$Canadian Coast Guard College, Dept.\ of Arts, Sciences, and Languages, Sydney, Nova Scotia, B1R 2J6, Canada\\
$^{6}$D\'epartement de physique and Centre de Recherche en Astrophysique du Qu\'ebec (CRAQ), Universit\'e de Montr\'eal, C.P. 6128, Succ.\ Centre-Ville, Montr\'eal, Qu\'ebec, H3C 3J7, Canada\\
$^7$Ritter Observatory, Department of Physics and Astronomy, University of Toledo, Toledo, OH 43606-3390, USA\\
$^8$Department of Physical Sciences, Kutztown University, Kutztown, PA 19530, USA\\
$^9$Barfold Observatory, Glenhope, Victoria 3444, Australia\\
$^{10}$SASER -- the Southern Astro Spectroscopy Email Ring\\
$^{11}$International Centre for Radio Astronomy Research, University of Western Australia, 35 Stirling Hwy, Crawley, WA 6009, Australia\\
$^{12}$Harvard-Smithsonian Center for Astrophysics, 60, Garden street, Cambridge MA 02138, USA\\
$^{13}$Instituut voor Sterrenkunde, Celestijnenlaan 200D, B-3001 Leuven, Belgium\\
$^{14}$Department of Physics, University of Antwerp, Groenenborgerlaan 171, 2020 Antwerp, Belgium\\
$^{15}$Department of Physics and Astronomy, University of Louisville, Louisville, KY 40292, USA\\
$^{16}$Department of Physics and Astronomy, Vanderbilt University, Nashville, TN 37235, USA\\
$^{17}$Department of Physics, Lehigh University, Bethlehem, PA 18015, USA\\
$^{18}$Department of Physics, Fisk University, 1000 17th Avenue North, Nashville, TN 37208, USA\\
$^{19}$Las Cumbres Obs.\ Global Telescope Network, 6740 Cortona Dr., Suite 102, Santa Barbara, CA 93117, USA\\
$^{20}$American Association of Variable Star Observers, Cambridge, MA 02138, USA\\
$^{21}$Center for Theoretical Physics of the Polish Academy of Sciences, Al. Lotnikow 32/46, 02-668 Warsaw, Poland\\
$^{22}$Faculty of Physics, University of Warsaw, Pasteura 5, 02-093 Warszawa, Poland\\
$^{23}$National Centre for Nuclear Research, Ho\.za 69, 00-681 Warsaw, Poland\\
$^{24}$International Centre for Radio Astronomy Research (ICRAR), Curtin University, GPO Box U1987, Perth, WA 6845, Australia\\
$^{25}$Institute for Computational Astrophysics, Department of Astronomy and Physics, Saint Mary's University, Halifax, NS B3H 3C3, Canada\\
$^{26}$Institut f\"ur Astronomie, Universit\"at Wien, T\"ur\-ken\-schanzstrasse 17, A-1180 Wien, Austria\\
$^{27}$Department of Astronomy and Astrophysics, University of Toronto, Toronto, ON M5S 3H4, Canada\\

\label{lastpage}

\end{document}